\documentclass[fleqn,usenatbib]{mnras}

\usepackage{newtxtext,newtxmath}

\usepackage[T1]{fontenc}
\usepackage{ae,aecompl}

\newcommand{\hMsun}{{\ifmmode{h^{-1}{\rm {M_{\odot}}}}\else{$h^{-1}{\rm{M_{\odot}}}$}\fi}}
\newcommand{\Msun}{{\ifmmode{{\rm {M_{\odot}}}}\else{${\rm{M_{\odot}}}$}\fi}}
\newcommand{\Tab}[1]{Table~\ref{#1}}
\newcommand{\Sec}[1]{Section~\ref{#1}}

\newcommand{\Fig}[1]{Fig.~\ref{#1}}
\newcommand{\App}[1]{Appendix~\ref{#1}}

\usepackage{graphicx}	        
\usepackage{amsmath}	        
\usepackage{amssymb}	        
\usepackage{multirow}           
\usepackage[dvipsnames]{xcolor} 

\newcommand{\code}[1]{\texttt{#1}}

\title[The evolution of cluster density profiles]{The Three Hundred Project: The evolution of galaxy cluster density profiles}

\author[Mostoghiu et. al]
       {Robert Mostoghiu,$^{1}$\thanks{robert.mostoghiu@uam.es}
        Alexander Knebe$,^{1,2,3}$
        Weiguang Cui,$^{1}$
        Frazer R. Pearce,$^{4}$
        \newauthor
        Gustavo Yepes,$^{1,2}$
        Chris Power,$^{3}$
        Romeel Dave,$^{5}$
        Alexander Arth$^{6,7}$
\\
\vspace{-.3cm}
$^{1}$Departamento de F\'isica Te\'{o}rica, M\'{o}dulo 15, Facultad de Ciencias, Universidad Aut\'{o}noma de Madrid, E-28049 Madrid, Spain\\ \vspace{-.3cm}
$^{2}$Centro de Investigaci\'{o}n Avanzada en F\'isica Fundamental (CIAFF), Facultad de Ciencias, Universidad Aut\'{o}noma de Madrid,\\\vspace{-.3cm} 28049 Madrid, Spain\\ \vspace{-.3cm}
$^{3}$International Centre for Radio Astronomy Research, University of Western Australia, 35 Stirling Highway, Crawley,\\\vspace{-.3cm} Western Australia 6009, Australia\\ \vspace{-.3cm}
$^{4}$School of Physics \& Astronomy, University of Nottingham, Nottingham NG7 2RD, UK\\ \vspace{-.3cm}
$^{5}$Institute for Astronomy, School of Physics \& Astronomy, The University of Edinburgh, Royal Observatory, Edinburgh EH9 3HJ, UK\\ \vspace{-.3cm}
$^{6}$University Observatory Munich, Scheinerstra{\ss}e 1, 81679 Munich, Germany\\ \vspace{-.3cm}
$^{7}$Max-Planck-Institute for Extraterrestrial Physics, Giessenbachstrasse 1, 85748 Garching, Germany\\ \vspace{-.3cm}
}

\date{Accepted XXX. Received YYY; in original form ZZZ}

\pubyear{2018}

\begin{document}
\label{firstpage}
\pagerange{\pageref{firstpage}--\pageref{lastpage}}
\maketitle

\begin{abstract}
Recent numerical studies of the dark matter density profiles of massive galaxy clusters ($M_{\rm halo}> 10^{15}$M$_{\odot}$) show that their median radial mass density profile remains unchanged up to $z>1$, displaying a highly self-similar evolution. We verify this by using the data set of the \textsc{The Three Hundred} project, i.e. 324 cluster-sized haloes as found in full physics hydrodynamical simulations. We track the progenitors of the mass-complete sample of clusters at $z=0$, and find that their median shape is already in place by $z=2.5$. However, selecting a dynamically relaxed subsample ($\sim16$ per cent of the clusters), we observe a shift of the scale radius $r_s$ towards larger values at earlier times. Classifying the whole sample by formation time, this evolution is understood as a result of a two-phase halo mass accretion process. Early-forming clusters -- identified as relaxed today -- have already entered their slow accretion phase, hence their mass growth occurs mostly at the outskirts. Late-forming clusters -- which are still unrelaxed today -- are in their fast accretion phase, thus the central region of the clusters is still growing. We conclude that the density profile of galaxy clusters shows a profound self-similarity out to redshifts $z\sim2.5$. This result holds for both gas and total density profiles when including baryonic physics, as reported here for two rather distinct sub-grid models.
\end{abstract}




\section{Introduction} \label{sec:introduction}
In the hierarchical growth paradigm of a $\Lambda$ cold dark matter ($\Lambda$CDM) Universe, where the continuous merging of lower mass systems into haloes yields more massive systems, galaxy clusters are the biggest gravitationally bound systems. Although they are dark-matter dominated, and thus their growth is driven by gravitational interaction, at the same time their properties are also determined by the interaction with the baryonic component of clusters. For this reason, galaxy clusters form ideal laboratories for understanding the underlying cosmology of our Universe and the physical processes driving galaxy evolution. 

It has been shown that for a flat Universe with scale-free initial density fluctuations, and at scales where baryonic physics can be neglected, dark matter haloes evolve self-similarly \citep[e.g.][]{Kaiser86}. And while the general shape of their density profiles is well defined by the two-parameter family of Navarro-Frenk-White (NFW) profiles \citep{Navarro96}, individually there can be great deviations --especially in the outer parts \citep{Diemer14}. Observationally, scaling relations between directly measurable cluster properties, such as the X-ray luminosity ($L_{X}$) or temperature ($kT$), can often be approximated by power laws. Hence, once the clusters are normalized by their mass, self-similar models can predict the slope of such relations \citep{Bower97}. However, non-gravitational processes \citep[e.g.][]{Pearce00, Voit03, Kravtsov06, Maughan12} are known to disrupt self-similarity and thus deviations are observed from such theoretical predictions, albeit observational results suggest that the influence of these processes is confined within the innermost regions of the clusters \citep[e.g.][]{Bartalucci17a, McDonald17, Ghirardini18}. Moreover, defining cluster masses and scales in terms of a characteristic density of the Universe (e.g. the background or critical density) induces an apparent evolution or ``pseudo-evolution'' on the cluster \citep{Bryan98,Diemer13}: the halo grows not only through dynamical processes, such as the merger of haloes, but also due to the change in the redshift-dependent reference density, which changes the normalization of the observed scaling relation. This needs to be borne in mind when interpreting results especially from numerical simulations where various mass definitions are being used throughout the literature \citep[see][for a discussion about possible halo mass definitions in simulations]{Knebe13b}.  

The growth of the universal dark matter halo density profile \citep[as found in simulations of cosmic structure formation,][]{Navarro97} is often described in terms of a two-phase process: an early fast accretion phase where mass builds up in the central region of the cluster, and a later slow accretion phase, where the mass builds in the outer region of the cluster while the mass density in the inner region remains approximately constant \citep{Gott75, Gunn77, Lokas00, Ascasibar04, Ascasibar07, Lithwick11}. These accretion modes are also observed in numerical simulations \citep[e.g.][]{Bullock01,Wechsler02,Zhao03,Diemer14}. Nevertheless, the numerical study of galaxy clusters at high redshift is a computationally-demanding task. In order to compare the results with observations, the profiles obtained from simulations are required to achieve spatial resolution down to the kpc scale \citep[e.g.][]{Bartalucci17a,Bartalucci17b, Ruppin17} in order to resolve the inner regions of the clusters where complex baryonic processes, like feedback from stars or active galactic nuclei (AGN), take place. At the same time, the simulations need to correctly reproduce the influence of the large-scale structure on the outer \--- gravity-dominated \--- regions of the profile.

In a recent article, \citealt{LeBrun18}, (from now on, LB+18) have shown that the 25 most massive galaxy clusters, as found in their dark matter only simulations and when scaled appropriately, have density profiles that are already in place at redshifts $z>1$ and hence remarkably robust to mergers and any other evolutionary effects. To this extent they have studied the median density profile of the most massive systems identified at the distinct redshifts $z=0, 0.6, 0.8$, and $1.0$. They did not track the progenitors of their $z=0$ sample and hence always had a mass complete set at every redshift studied. They found that the median density profiles of their sample of objects exhibits very little evolution with only a mild scatter of 0.15 dex.

In order to study the evolution of density profiles we have taken a different approach and followed the merging history of our sample at redshift $z=0$ backwards in time. Our data set stems from \textsc{The Three Hundred} project, i.e. a sample of over 300 galaxy clusters simulated with full-physics hydrodynamics \citep[see][for more details]{Cui18}. This sample includes within it a set of objects similar in mass and selection to that presented by LB+18, but all our clusters are simulated with the relevant baryonic physics as modeled by two different hydrodynamics solvers and full-physics sub-grid models alongside the same gravity solver (i.e. \code{GADGET-MUSIC} and \code{GADGET-X}, as described below). While we also have at our disposal the corresponding dark matter only simulations, here we only present results from the full-physics runs (see \App{app:most_massive} for a more direct comparison to the work of LB+18, and \App{app:dm_only} for the dark matter only results).

This paper is structured as follows: in Section \ref{sec:simulations} we present the database and the codes used in this analysis. In Section \ref{sec:cluster_sample} we describe the properties of our sample in detail. The results for the progenitors of the redshift $z=0$ sample are presented in Section \ref{sec:results}, classified by their dynamical state (Section \ref{sec:subsubsec:results_relaxed}) or by their formation time (Section \ref{sec:subsubsec:results_formationtime}). In Section \ref{sec:subsec:results_gasprofiles} we focus on the observational predictions from the redshift evolution of the (total) mass density profiles of the clusters by analyzing their corresponding gas mass density profiles from the complete sample. In Section \ref{sec:subsec:present_scales} we analyze the present day cluster scales in terms of their dynamical state and formation time. Finally, we conclude our analysis in Section \ref{sec:conclusions}.

\section{Simulation and codes} \label{sec:simulations}

The clusters within the \textsc{The Three Hundred} dataset were created by extracting 324 spherical regions of $15 h^{-1}$ Mpc radius centered on each of the most massive clusters identified at $z=0$ within the dark-matter-only MDPL2, MultiDark simulation \citep{Klypin16}.\footnote{The MultiDark simulations are publicly available at the \url{https://www.cosmosim.org} database.} The cosmological parameters of the MDPL2 simulation are based on the Planck 2015 cosmology \citep{Planck2015}, with $\Omega_{\rm M} = 0.307$, $\Omega_{\rm b} = 0.048$, $\Omega_\Lambda = 0.693$, $h=0.678$, $\sigma_8 = 0.823$, and $n_s=0.96$. The MDPL2 box has a side-length of $1 h^{-1}$ Gpc and it contains $3840^3$ dark matter particles each of mass $1.5 \times 10^9 h^{-1}$ M$_{\odot}$. The Plummer equivalent softening of the simulation is $6.5h^{-1}$ kpc.

To produce full-physics hydrodynamics simulations, multiple levels of mass refinement for the initial conditions have been generated using the fully parallel \code{GINNUNGAGAP}\footnote{\url{https://github.com/ginnungagapgroup/ginnungagap}} code. Particles that end up within the $15 h^{-1}$ Mpc radius spherical region at $z=0$ are traced back to the initial conditions and within these regions the corresponding dark matter particles are split into dark matter and gas particles according to the cosmological baryon fraction from the Planck 2015 cosmology, $\Omega_{\rm b}/\Omega_{\rm M} \sim 0.16$. Consequently, the simulations reach an effective dark matter and gas particle resolution of $m_{\rm DM}=1.27\times 10^{9} h^{-1}$ M$_{\odot}$ and $m_{\rm gas}=2.36\times 10^{8} h^{-1}$ M$_{\odot}$, respectively. To reduce the computational cost of the original simulation, outside the re-simulated regions, dark matter particles are successively degraded with lower mass resolution particles in order to maintain the same large scale tidal field.

We present here the results from running the initial conditions with two distinct codes, i.e. \code{GADGET-MUSIC} and \code{GADGET-X}: 

\code{GADGET-MUSIC} is a modified version of \code{GADGET3} (an updated version of \code{GADGET2} by \citealt{springel_gadget2_2005}) based on the classical entropy-conserving SPH formulation \citep{Springel02}, with a 40 neighbour M3 interpolation kernel. It features metallicity-independent radiative cooling, metal enrichment of gas and stars coming from supernovae, the effects of an ionizing UV homogeneous background based on \citet{Haardt01}, and a star formation model from \citet{Springel03}. However, it lacks a model for supermassive black hole growth or AGN feedback.

\code{GADGET-X} is also a modified version of \code{GADGET3}, with a 200 neighbour high-order Wendland C4 interpolation kernel. It features an improved and more modern SPH scheme \citep{Beck16} that gives a better description of discontinuities and clumpiness instabilities within the gas. The code includes an uniform UV background from \citet{Haardt96}, gas cooling following \citet{Wiersma09}, a star formation model with chemical enrichment from \citet{Tornatore07}, and kinetic supernova feedback and AGN feedback from \citet{Springel03} and \citet{Steinborn15}, respectively.

For the analysis of the haloes we used the \code{AHF}\footnote{\url{http://popia.ft.uam.es/AHF}} halo finder \citep{Gill04a,Knollmann09}, which locates local overdensities in an adaptively smoothed density field as potential halo centers and automatically identifies haloes and substructure (subhaloes, subsubhaloes, etc.). For every halo found, \code{AHF} calculates its radius $r_{200}$ (and corresponding enclosed mass $M_{200}$) as the radius $r$ at which the density $\rho(r)=M(<r)/(4\pi r^{3}/3)$ drops below $200\rho_{\rm crit}$, where $\rho_{\rm crit}$ is the critical density of the Universe at a given redshift $z$. Note, $r_{500}$ is defined accordingly.  Furthermore, \code{AHF} creates a number of radial profiles in log-spaced spherical shells out to $r_{200}$: on average, a cluster profile calculated by \code{AHF} at $z=0$ consists of about 35 logarithmically spaced radial bins covering the range $\sim 0.2-0.3$ $h^{-1}$kpc to $r_{200}\sim 1500$ $h^{-1}$kpc. Note that, due to the hierarchical way in which \code{AHF} identifies haloes from density peaks, the masses of the haloes found are inclusive, i.e. the mass of a host halo includes the mass of all of the subhaloes contained within that halo.

Finally, to trace haloes through the snapshots we build merger trees with \code{MergerTree}, a tool that comes with \code{AHF}. \code{MergerTree} has been used to follow each halo identified at redshift $z = 0$ backwards in time, identifying as the main progenitor at some previous redshift the halo that maximizes the merit function $\mathcal{M} =N_{A\cap B}^2/(N_{A} N_{B})$ where $N_A$ and $N_B$ are the number of particles in haloes $H_A$ and $H_B$, respectively, and $N_{A\cap B}$ is the number of particles that are in both $H_A$ and $H_B$. More details about the merit function and the performance of \code{MergerTree} can be found in \citet{Srisawat13}.

\section{Cluster sample} \label{sec:cluster_sample}

The study of the fundamental galaxy cluster properties and scaling relations of the sample can be found in the introductory paper of The Three Hundred Project \citep{Cui18}. Overall, the modelled galaxy clusters in both codes are in reasonable agreement with observations with respect to baryonic fractions and gas scaling relations at redshift $z=0$, with some (mode-dependent) differences, such as the existence of too massive central galaxies, or bluer galaxy colours (about 0.2 dex lower at the peak position) compared with observations.

With a simulated volume of radius $15 h^{-1}$ Mpc each of our regions contains many more objects than the large cluster located at the centre. However, in this paper we confine ourselves solely to considering the evolution of the central object, i.e. the object on which the spherical region is centred and that was originally chosen to be modeled. By construction, these objects form, at redshift $z=0$, a mass-complete sample of the largest objects in the full MDPL2 box \citep[see Fig. A1 in][]{Cui18}. But their progenitors will {\it not} form a mass-complete sample at higher redshifts. 

To quantify the dynamical state of our objects, we adopt the same estimators as presented in \citet{Cui18}. This means, that for each of our 324 large haloes we calculate three proxy indicators for virialization. The first parameter is the fraction of mass in subhaloes $f_s = \sum{M_{\rm sub}/M_{200}}$, where $M_{\rm sub}$ is the mass of each subhalo. The second parameter is the virial equilibrium parameter $\eta = (2T -E_{\rm s})/|W|$, where $T$ is the total kinetic energy, $E_{\rm s}$ the energy associated to the surface pressure exerted on the halo at $r_{200}$ due to in-falling material, and $W$ is the total potential energy. Finally, the last parameter is the center-of-mass offset $\Delta_{\rm r}=|r_{\rm cm} - r_{c}|/r_{200}$, where $r_{\rm cm}$ is the center-of-mass within a cluster radius $r_{200}$, and $r_{\rm c}$ is the center of the cluster corresponding to the maximum density peak of the halo. As described in \citet{Cui18}, the criteria for selecting dynamically relaxed clusters are: $0.85 < \eta < 1.15$, $\Delta_{\rm r} < 0.04$, and $f_{\rm s} < 0.1$, which need to be satisfied simultaneously \citep[see][for similar definitions]{Neto07,Power12,Cui17}. With these parameters, $\sim16$ per cent ($\sim17$ per cent for \code{G-X}; and $\sim15$ per cent for \code{G-MUSIC}) of the mass-complete sample are relaxed at $z=0$.

For the analysis of the density profiles we restrict our data to radii selected according to the following method. We first analyze the radial profiles of the MDPL2 clusters, avoiding inner bins where two-body collisions dominate the interaction between particles by requiring an estimate for the local collisional relaxation time to be larger than the age of the universe \citep{Power03}. As this convergence criterion is based upon dark matter only simulations, we have used the counterpart MDPL2 clusters to determine their maximum innermost converged radius $r^{\rm MDPL2}_{\rm conv}$ (usually determined by the least massive object in the sample), and guided by this radius, we selected the radial values from the hydrodynamical runs entering our analysis. We refer to this inner limit as the `validation' radius $r_{\rm valid}$: this turns out to be $r_{\rm valid}=r^{\rm MDPL2}_{\rm conv}\sim 28$ $h^{-1}$kpc (approximately 4 times the softening of the simulation) at $z=0$, and $\sim 37$ $h^{-1}$kpc, $\sim 44$ $h^{-1}$kpc, and $\sim 55$ $h^{-1}$kpc (in comoving coordinates) at redshifts $z=0.5,1$, and $2.5$, respectively. For each halo we focused our study on the mass density profile. While we mainly analyzed the total (dark and baryonic matter) mass density profiles, we also studied the dark-matter-only and gas profiles from the hydrodynamical simulations (and the dark-matter-only profiles from the underlying dark-matter-only simulation, see \App{app:dm_only}); none of that has an effect on the results and conclusions drawn from them, respectively, and hence we decided to only show the total mass results here.

By using two hydrodynamical codes with the same gravitational treatment but different SPH recipes and modelled baryonic physics, we are able to study the influence \--- if any \--- of these changes on the evolution of the density profiles investigated in detail in the following Section. However, we expect the influence of different baryonic processes to be (mostly) reflected in the innermost regions of the clusters, where we find pronounced deviations from a Navarro-Frenk-White (NFW) profile \citep{Navarro96} that otherwise describes our cluster profiles remarkably well. 

\begin{table}
\centering
\caption{Minimum, median, and maximum mass values of the 324 cluster mass-complete sample at $z=0$ sample and their progenitors at each redshift (see text for further details). In each row, the left value corresponds to the \code{GADGET-X} simulation, the right one to \code{GADGET-MUSIC}. All values are in units of $10^{14}h^{-1}$M$_{\odot}$.}
\label{tab:stat_values_halo_mass_progenitorselect}
\resizebox{\columnwidth}{!}{
\begin{tabular}{lcccccc}
\hline
\hline
Redshift & \multicolumn{2}{c}{min($M_{200}$)}    & \multicolumn{2}{c}{med($M_{200}$)}   & \multicolumn{2}{c}{max($M_{200}$) }\\
   & \code{G-X} & \code{G-MUSIC} & \code{G-X} & \code{G-MUSIC} & \code{G-X} & \code{G-MUSIC}\\
\hline
$z=0$      & 5.22 & 5.18                           & 8.22 & 8.27                        & 26.21 & 26.22 \\ 
\hline
$z=0.5$    & 0.58 & 0.31                           & 3.90 & 3.80                        & 18.93 & 18.95 \\ 
\hline
$z=1$      & 0.23 & 0.21                           & 1.64 & 1.73                        & 6.90  & 6.89  \\ 
\hline
$z=2.5$   & 0.0071 & 0.0065                        & 0.16 & 0.17                        & 1.77  & 1.80  \\ 
\hline
\hline
\end{tabular}
}
\end{table}

\section{Results} \label{sec:results}

From the mass density profiles obtained for each galaxy cluster in the sample, we calculated the median  profile at each redshift. To calculate the median profiles we first defined 35 radial bins between the minimum and the maximum radial bin of the whole sample. We then interpolate each individual profile to those radii. If --for a given cluster-- the profile would require extrapolation to one of our predefined radial bins, we instead flag that radial bin indicating that not all clusters can contribute to it. The resulting median values at our 35 radii are calculated using (non-flagged) interpolated values requiring contribution from at least 50 per cent of the sample. This criterion reduces the number of bins of the median profiles to $\sim30$ bins, with inner and outer limits of $\sim 6 h^{-1}$kpc and $\sim 1274 h^{-1}$kpc at $z=0$, respectively. Although we are not interested in the innermost region of the profile for this study, we also flag the unvalidated values (i.e. $r<28,37,44,$ or $55 h^{-1}$kpc depending on the redshift) calculated from at least 50 per cent of the clusters, i.e. inner profile values which pass the first criterion but fail the second (for more details about the median profiles, see \App{app:prof_properties}).

Following the work of LB+18, the density profiles have been normalized by the critical density of the Universe at the corresponding redshift, and the radii have been scaled by $r_{500}$. When plotting the density profiles we further multiplied them by $(r/r_{500})^2$ to reduce the dynamical range. This also allows us to determine the scale radius $r_{s}$ via the peak position $x_{\rm peak}$ of the resulting curve\footnote{Formally speaking, by multiplying the density profile with $r^2$, the peak position is characterized by $r_{-2}$, i.e. the position where the logarithmic slope is $-2$. But for a NFW profile $r_s=r_{-2}$ and hence we refer to it as the scale radius.} 
\begin{equation} \label{eq:scaleradius}
 r_s = x_{\rm peak} \ r_{500} \ ,
\end{equation}
where the median $r_{500}$ value will be used. The peak position of a median profile is found by first selecting  the values found in the region $r/r_{500}>0.2$ and $r/r_{500}<1.1$ at our 4 redshift. The median profile values in the selected region have then been interpolated at 1000 points on this interval using 3rd order splines. From the interpolated profiles we then find the (scaled) radii at which the profiles reach their maximum. To obtain the uncertainty of the peak we started from the $30-70^{\rm th}$ percentiles of the median profile at a redshift value. We then applied the same procedure we used for the median profile to both percentiles curves and obtained an uncertainty interval for each peak's position (i.e. selecting a region, interpolating the original data, and finding the maximum's position). Next, we apply a Savitzky-Golay filter \citep{Savitzky1964} with a 5-point window and a 3rd order spline to the interpolated data. Note that we use a relatively small window for the smoothing. This is done in order to ensure that  the high redshift median profiles are being smoothed.\footnote{After removing the innermost bins $r<r_{\rm valid}$ (see explanation in \Sec{sec:cluster_sample}), ensuring that at least half the sample enters the median calculation, and selecting the region for the interpolation, we might end up with $\sim7$ points at the interpolation stage.} Using the smoothed interpolated data, we find the maximum's position.

According to LB+18, we should not expect to find any evolution out to redshift $z\sim1$. We note that the result from LB+18 was obtained for mass-complete samples at the corresponding redshifts. However, we are following the central haloes found at redshift $z=0$ backwards in time. We therefore consider it relevant to summarize in \Tab{tab:stat_values_halo_mass_progenitorselect} the minimum, median, and maximum masses of our objects at all redshifts of relevance. It is worth mentioning that the central halo from region `0047' is undergoing merger event at $z=0$. During such events, halo finders are known to present some problems (see, for instance, Fig.4 in \citealt{Behroozi15}). We found that for \code{G-X}'s central halo `0047' \code{AHF} assigned the other cluster participating in the merger as the host halo; hence obtaining $\sim20$ per cent less mass than in its \code{G-MUSIC} counterpart. Thus, we removed this central halo from the results in \Tab{tab:stat_values_halo_mass_progenitorselect}.

\subsection{Dependence on Dynamical State}\label{sec:subsubsec:results_relaxed}
\begin{figure*}
	\includegraphics[width=1\textwidth]{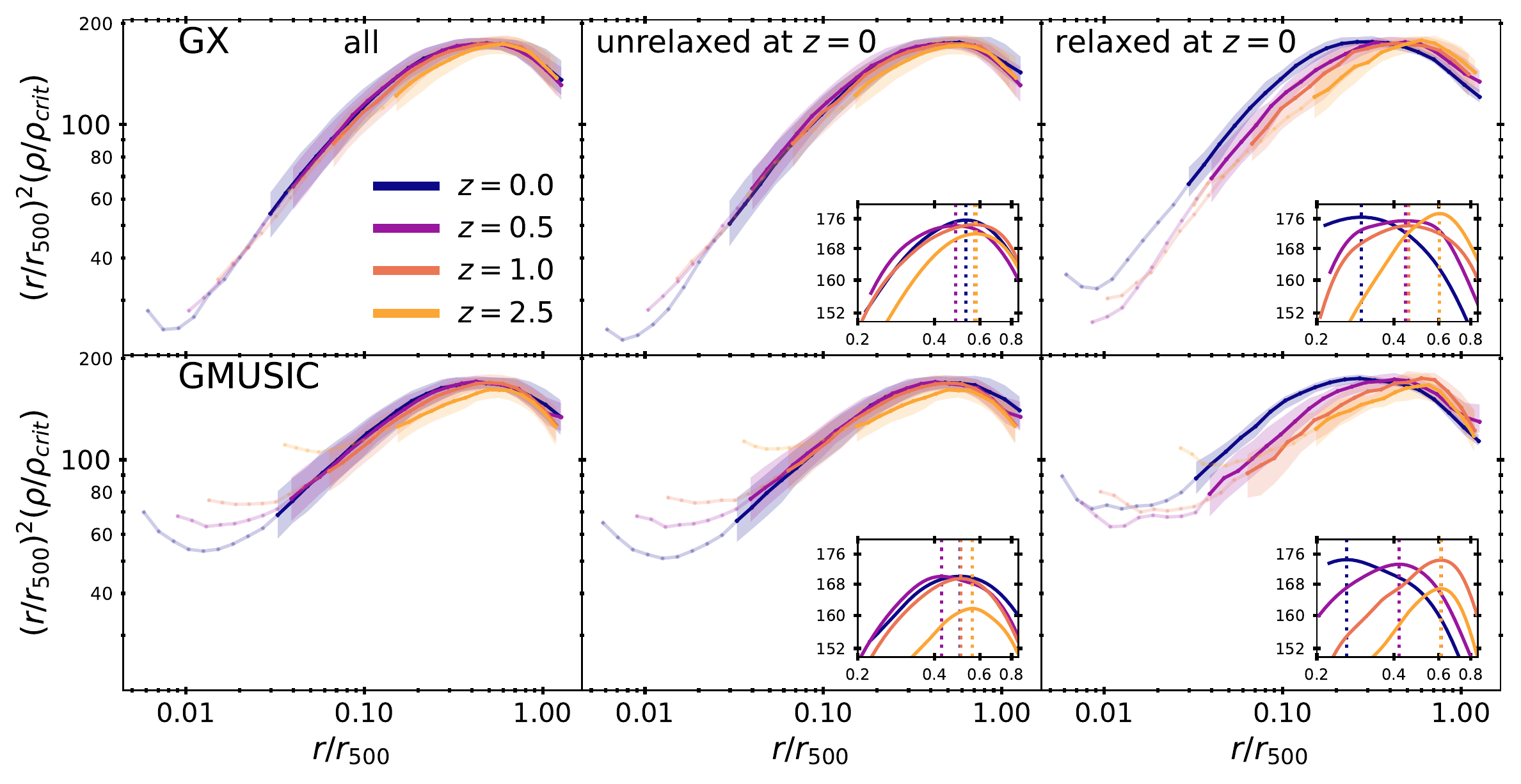}
    \caption{Median scaled mass density profiles of the central haloes at $z=0$ and their main progenitors at $z=0.5$, $1$, and $2.5$. The top row shows the results from \code{GADGET-X}, the bottom row from \code{GADGET-MUSIC}. The first column shows the results for the whole mass-complete sample (324 clusters), while the unrelaxed ($\sim84$ per cent of the total sample) and the results from the relaxed ($\sim16$ per cent) subsamples are presented in the second and third columns, respectively. The shadowed regions represent the $30-70$ percentiles. Unvalidated values are shown in lighter colors. In the insets we show the peak's position, $r_s/r_{500}$, for every curve. Overall, the height of the density peak remains constant (within the errors) for all the samples. As seen in the insets, we detect a clear shift in the density peak's position in the relaxed subsample.}
    \label{fig:density_median_relaxation}
\end{figure*}

Besides showing the evolution of the density profile in the left column of \Fig{fig:density_median_relaxation}, we also split our sample of objects according to their dynamical state at redshift $z=0$. The results are shown in the middle column for the unrelaxed clusters ($\sim84$ per cent of the total sample), and in the right column for the relaxed clusters ($\sim16$ per cent). The two different rows correspond to  \code{G-X} (top) and \code{G-MUSIC} (bottom). The $30-70$ percentiles are represented by the shadowed regions. Unvalidated values calculated from at least 50 per cent of the clusters are shown in lighter colors. Besides deviations attributed to baryonic physics in the inner regions of the profiles and the expected influence from different environments at the outermost regions, the profiles present a strikingly self-similar evolution within $0.1<r/r_{500}<1$, with only a slight deviation in the maximum position at different redshifts. This is in agreement with previous results found by LB+18: after rescaling the density we find that the density profiles of the massive haloes of galaxy clusters evolve self-similarly.

However, when classified by their dynamical state, we observe some remarkable differences: unlike the complete sample, for the relaxed clusters there is a clear shift in the position of the maximum $r_{s}/r_{500}$ with redshift, decreasing as we move to smaller redshift. To better identify the shift, we added insets for the subsamples of the total 324 central haloes sample. In the insets of \Fig{fig:density_median_relaxation} we can see that, while the whole sample presented a maximum at $r_{s}/r_{500}\sim0.6$, once we select relaxed objects we see that they display a shift in the maximum position from redshift $z=2.5$ onwards, reaching $r_{s}/r_{500}\sim0.3$ at $z=0$. While we primarily studied the density peak's position shift, note that, qualitatively, the peak value $(r_{\rm s}/r_{500})^2(\rho_{\rm s}/\rho_{\rm crit})$ is approximately constant (within the errors), where $\rho_{\rm s}$ is the density of the halo at $r=r_{\rm s}$. 

As we observe a shift in the ratio $r_s/r_{500}$, we now explore in more detail whether this is caused by a decrease in $r_s$, an increase in $r_{500}$, or a combination of both. To this extent we determine the peak positions as seen in \Fig{fig:density_median_relaxation} via spline-interpolation and the median $r_{500}$ for the respective sample. This allows us to calculate the corresponding $r_s$ via Eq.~(\ref{eq:scaleradius}). Another approach to retrieve these numbers would be to use $r_{\rm s}$ and $r_{500}$ values obtained from individual fits of the density profile to the functional form of an NFW profile. We confirm that this does not alter or affect our results.

\begin{figure*}
	\includegraphics[width=1\textwidth]{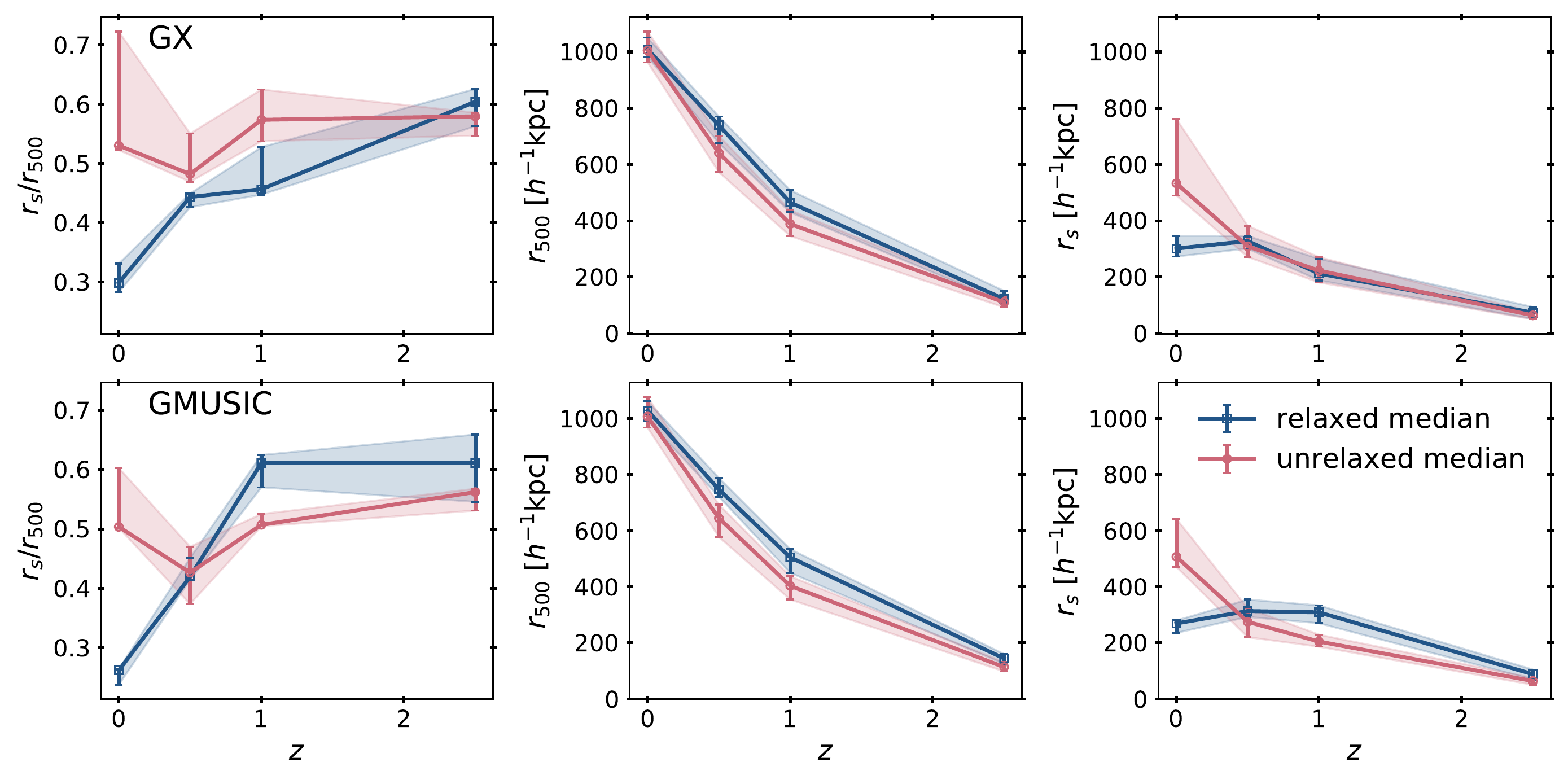}
    \caption{Redshift evolution of the density peak's position $r_{\rm s}/r_{500}$ (first column), $r_{500}$ (second column) and the scale radius, $r_{s}$ (third column), all in physical units, for the central cluster haloes at redshift $z=0$ and their main progenitors at redshift $z=0.5$, $1$, and $2.5$, classified by their dynamical state at redshift $z=0$ ($\sim84$ per cent unrelaxed, $\sim16$ per cent relaxed). In the top row, we show the results from \code{GADGET-X}; in the bottom row, for \code{GADGET-MUSIC}. While $r_{500}$ is a combined result of halo accretion and pseudo-evolution, the evolution of $r_{\rm s}$ suggests that the relaxed and unrelaxed subsamples are experiencing different accretion phases as $z=0$ is approached.}
    \label{fig:radii_evolution_relaxation}
\end{figure*}

The redshift evolution of the ratio $r_{\rm s}/r_{500}$ is presented in Fig \ref{fig:radii_evolution_relaxation} (left column), alongside the individual evolution of $r_{500}$ (middle column) and $r_{\rm s}$ (right column) -- all in physical units.  The upper row shows results for \code{G-X} whereas the lower row shows \code{G-MUSIC}. The relaxed and unrelaxed samples are color-coded according to the legend. As we can see, $r_{500}$ grows monotonically down to $z=0$ for both the relaxed and unrelaxed subsamples, with a slightly steeper slope in the unrelaxed subsample. The growth of $r_{500}$ is a result of the combined effect of accretion and pseudo-evolution, however, pseudo-evolution affects both subsamples equally. On the other hand, we see that $r_{s}$ undergoes a different evolution depending on the dynamical state of the clusters. For unrelaxed clusters, $r_{s}$ grows monotonically down to $z=0$ with a similar slope to the observed growth in the physical $r_{500}$, thus the overall evolution of the density peak is not seen in the median trend. For the relaxed clusters, we see an initial period of growth, similar to the one in the unrelaxed subsample, until at $z\approx 0.5$ the growth slows down and remains close to a constant value. We can understand this difference as a result of the different accretion phases each subsample experiences close to $z=0$, something expected from the secondary infall model \citep[e.g.][]{Ascasibar04, Ascasibar07}.

\subsection{Dependence on Formation Time}\label{sec:subsubsec:results_formationtime}

As the dynamical state of a galaxy cluster is linked to its formation time (i.e. earlier formed systems have had more time to relax and eventually pass from the phase of early, fast accretion to the stage of late, slow accretion), we now sub-divide our mass-complete sample at redshift $z=0$ into several formation redshift bins as follows: to determine the formation time we fit the mass accretion history (MAH) of each cluster to the functional form proposed in \citet{Wechsler02}.
\begin{equation}
M_{200}(z) = M_{200}^{z=0} \ \ \exp{(-\alpha z)},
\end{equation}

\noindent
where $\alpha$ is a characteristic parameter for each cluster which describes the (assumed constant) mass accretion rate of the halo. The formation redshift $z_{\rm form}$ is now defined as the redshift where $M_{200}(z_{\rm form})/M_{200}^{z=0}=0.5$ or equivalently
\begin{equation}
z_{\rm form}= -\ln{(0.5)}/\alpha .
\end{equation}

\begin{figure*}
	\includegraphics[width=1\textwidth]{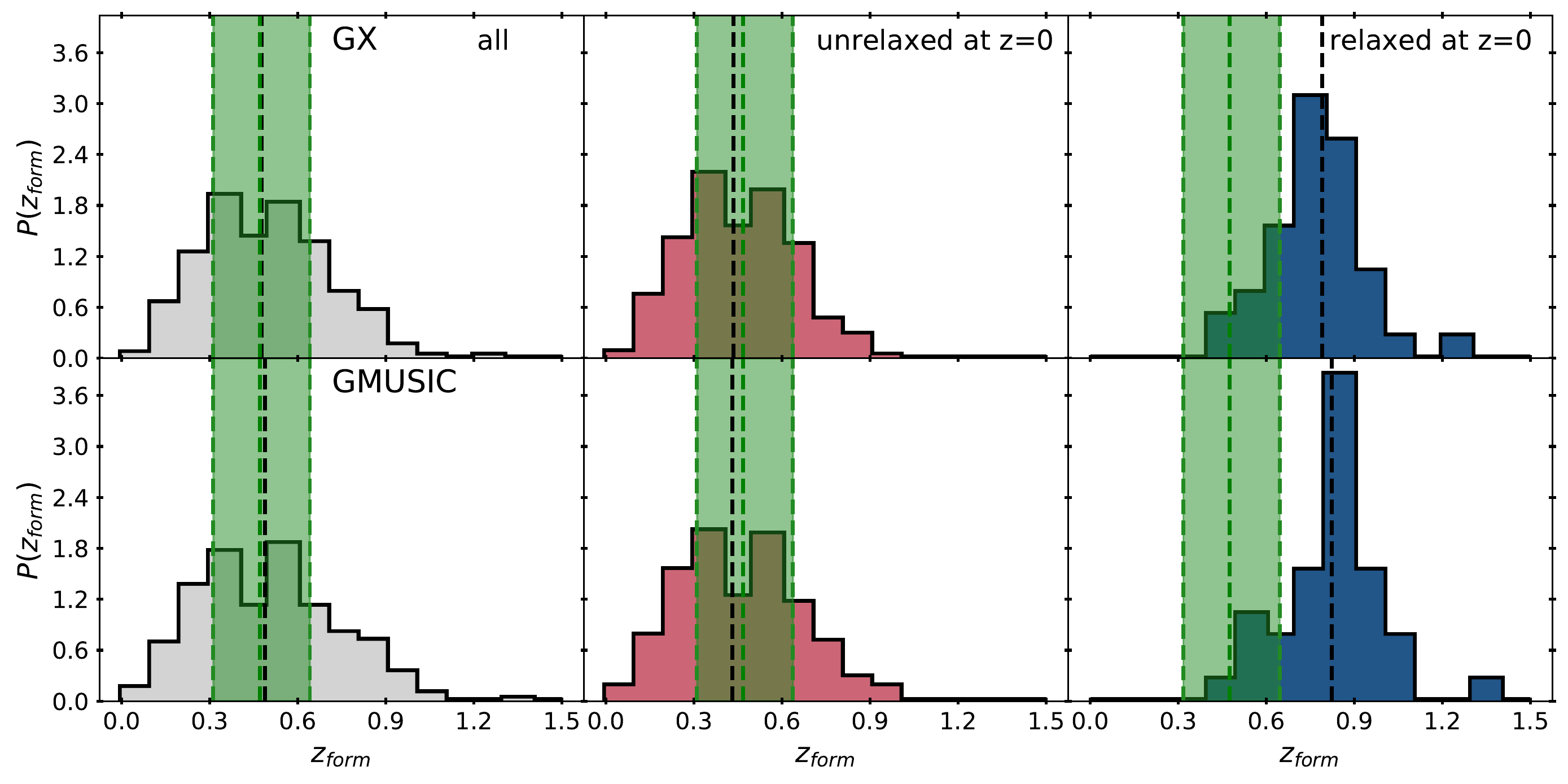}
    \caption{Formation time distribution of the whole mass-complete sample (left column), the unrelaxed (middle column) and the relaxed (right column) at $z=0$ subsamples, as obtained from \code{GADGET-X} (top row) and \code{GADGET-MUSIC} (bottom row). The black dashed line represents the median formation time of each sample. The middle green dashed line and green shaded regions show the expected formation time and $1\sigma$ errors, respectively, from the extended Press-Schechter calculation of \citet{Power12}, for the median halo mass of each sample. While both the complete sample and the unrelaxed subsample formed within the expected values ($z_{\rm form}\sim0.5$ and $z_{\rm form}\sim0.4$, respectively), the relaxed subsample formed earlier, at $z_{\rm form}\sim0.8$. The total mass density profile of the clusters is already in place at $z=2.5$, long before their formation time.}.
    \label{fig:formation_time_distribution_relaxation}
\end{figure*}

\noindent
With this definition, in Fig. \ref{fig:formation_time_distribution_relaxation} we show the distribution of formation redshift of the whole mass-complete sample (first column), the unrelaxed subsample (second column), and the relaxed subsample (third column) for the two hydrodynamical simulations in our database, \code{G-X} (top row), and \code{G-MUSIC} (bottom row). The dashed black line shows the median formation time of each sample, and the green dashed line and green shaded region show the expected median formation time and 1$\sigma$ errors from the extended Press-Schechter theory calculation of \citet{Power12}. The median formation time of the mass-complete sample is in good agreement with the expected value for haloes of the same median mass. However, although there is an overlap in the formation time distributions of the unrelaxed and relaxed subsamples, i.e. we find both unrelaxed and relaxed clusters within $0.4\leq z_{\rm form}\leq 1$, the relaxed clusters formed well before the expected value. This confirms the correlation between formation time and dynamical state \citep[e.g.][]{Power12,Wong12}. Note that -- according to the results presented in \Sec{sec:subsubsec:results_relaxed} -- the total mass density profile of the clusters is found to be already in place at $z=2.5$, long before their formation time.

\begin{figure*}
	\includegraphics[width=1\textwidth]{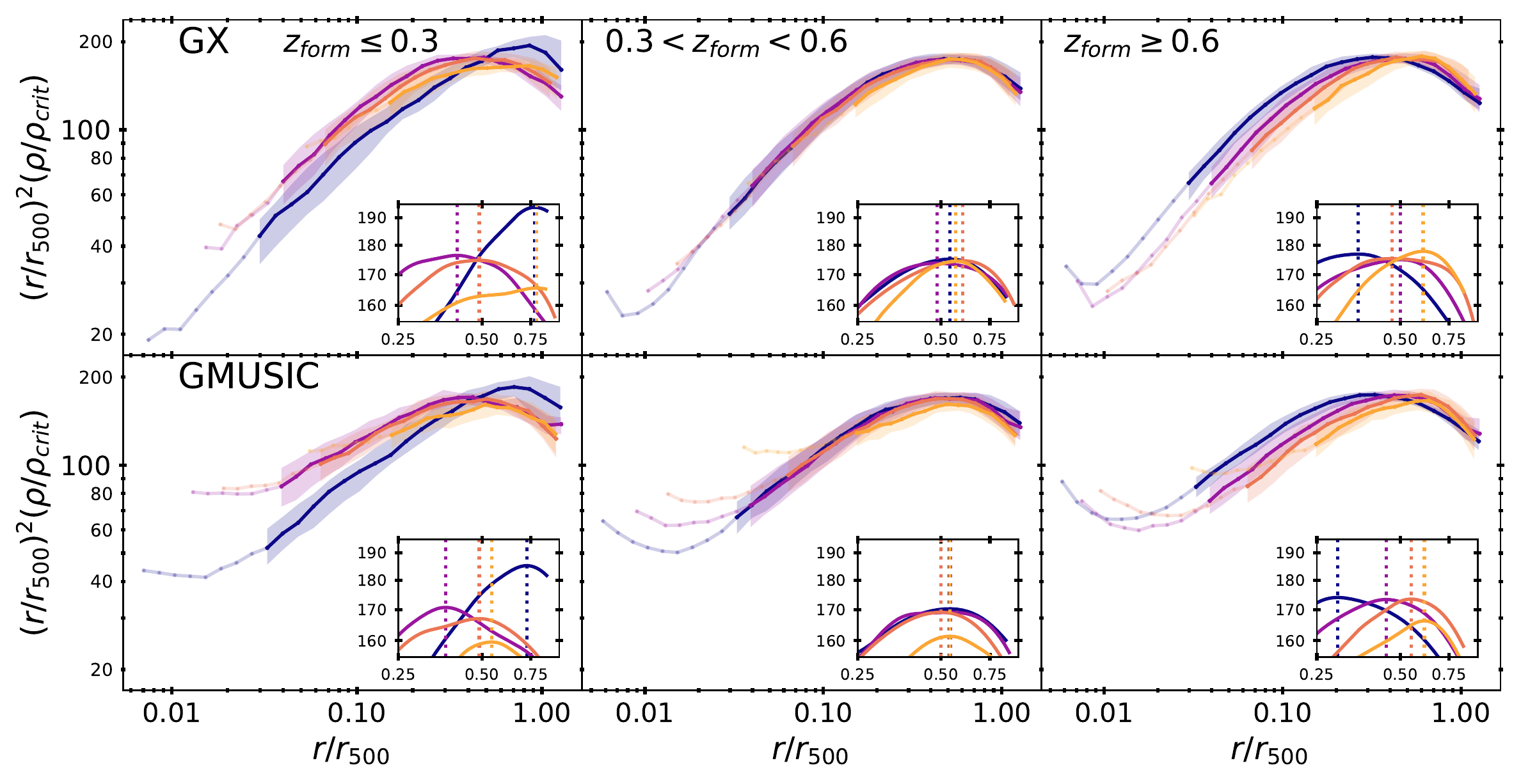}
    \caption{Median scaled mass density profiles of the central haloes at $z=0$ and their main progenitors at $z=0.5$, $1$, and $2.5$, classified by their formation time. The top row shows the results from \code{GADGET-X}, the bottom row \code{GADGET-MUSIC}. Unvalidated values are shown in lighter colors. The sample was separated into late-formed ($z_{\rm form}\leq 0.3$, $\sim21$ per cent of the total sample, left column), intermediate-formed ($0.3<z_{\rm form}< 0.6$, $\sim49$ per cent, middle column), and early-formed ($z_{\rm form}\geq 0.6$, $\sim30$ per cent, right-column) clusters, as found in both simulations. Similarly to what we found for a dynamical state classification, we detect a clear shift in the density peak's position in the early-formed subsample.}
    \label{fig:density_median_formtime}
\end{figure*}

\begin{figure*}
	\includegraphics[width=1\textwidth]{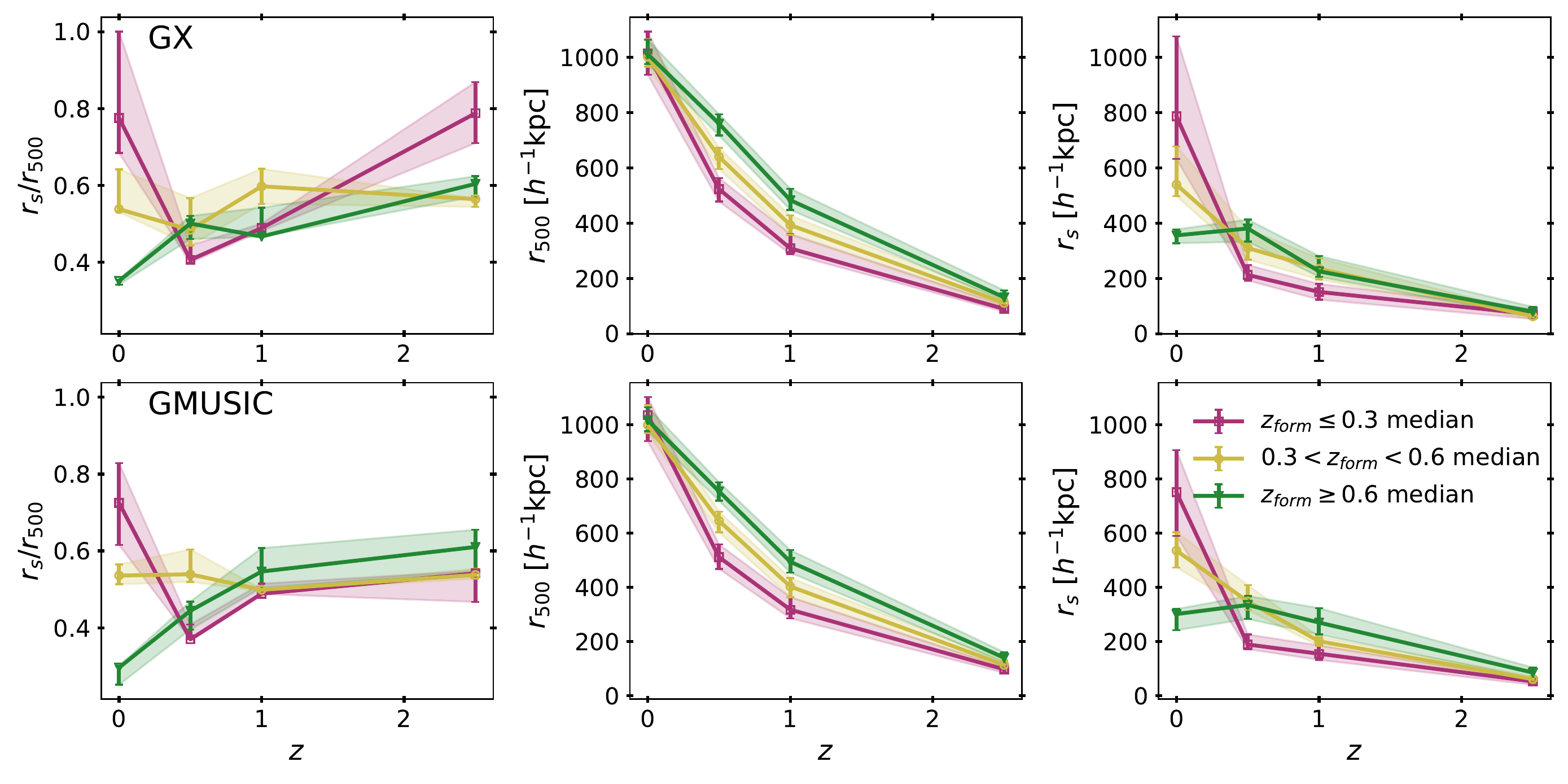}
    \caption{Redshift evolution of the density peak's position $r_{\rm s}/r_{500}$ (first column), $r_{500}$ (second column) and the physical scale radius, $r_{s}$ (third column), for the central cluster haloes at redshift $z=0$ and their main progenitors at redshift $z=0.5$, $1$, and $2.5$, classified by their formation time. In the top row, we show the results from \code{GADGET-X}; in the bottom row, for \code{GADGET-MUSIC}. The sample was separated into late-formed ($z_{\rm form}\leq 0.3$, $\sim21$ per cent of the total sample, left column), intermediate-formed ($0.3<z_{\rm form}< 0.6$, $\sim49$ per cent, middle column), and early-formed ($z_{\rm form}\geq 0.6$, $\sim30$ per cent, right-column) clusters. We observe a similar trend for early-formed haloes as in the dynamically relaxed subsample in \Fig{fig:radii_evolution_relaxation}.}
    \label{fig:radii_evolution_formtime}
\end{figure*}

To confirm the link between the formation time of our clusters and the shift observed in the peak position for $r_s/r_{500}$ for dynamically relaxed clusters at $z=0$, we present in \Fig{fig:density_median_formtime}  the evolution of the median density profile again, but this time for three distinct formation time bins chosen to minimize overlap in formation times and also to give roughly equal numbers of clusters at the two extremes in both simulations (more on this later), i.e. $z_{\rm form}<0.3$ ($\sim21$ per cent of the total sample, left column), $0.3<z_{\rm form}<0.6$ ($\sim49$ per cent, middle column), and $z_{\rm form}>0.6$ ($\sim30$ per cent, right column); the two rows are again for \code{G-X} (upper row) and \code{G-MUSIC} (lower row). The plot confirms that early formed system show a shift, but also that there is substantial scatter for late formed objects. \Fig{fig:radii_evolution_formtime} further quantifies the shift in $r_s$, $r_{500}$, and its ratio. This plot is analogous to \Fig{fig:radii_evolution_relaxation}, but now there are three lines in each panel, one for each redshift of formation bin. We obtain a similar shift once we turn to the early formed ($z\geq0.6$) subsample. Comparing the early-formed (green) and late-formed (pink) subsamples within each code, we find that the $r_{500}$ and $r_{\rm s}$ growth is similar to what we observed in the evolution of the dynamically relaxed and unrelaxed clusters in \Fig{fig:radii_evolution_relaxation}. Between $z=2.5$ and $z=1$, the early-formed clusters grow faster than the late-formed ones; however, between $z=1$ and $z=0.5$, both cluster subsamples appear to have approximately the same $r_{500}$ growth rates in both codes. Although the $r_{\rm s}$ growth rate of \code{G-X}'s early-formed clusters is steeper than \code{G-MUSIC}'s, there are no major differences in terms of $r_{\rm s}/r_{500}$. It is not until redshift $z<0.5$ that we observe a considerable difference in the peak's position evolution. At $z=0$ we end up with a difference of $\Delta(r_{\rm s}/r_{500})\sim0.5$ between the early and late-formed clusters. Similar to the trends observed in \Fig{fig:radii_evolution_relaxation}, the $r_{500}$ growth rate at $z<0.5$ of the late-formed clusters becomes much steeper than the rate of the early-formed, such that both subs-samples end up with a similar size at $z=0$. The $r_s$ evolution, however, differs. Unlike the late-formed clusters, the median $r_{\rm s}$ of the early-formed clusters remains constant from $z=0.5$. This result confirms that the formation time of the clusters drives the shift observed in the relaxed subsample and that the dynamical state is not the primary driver of it, even though there is a correlation between early-formed clusters and relaxed clusters, as seen from the formation time distribution presented in \Fig{fig:formation_time_distribution_relaxation}. 

\begin{figure}
	\includegraphics[width=1\columnwidth]{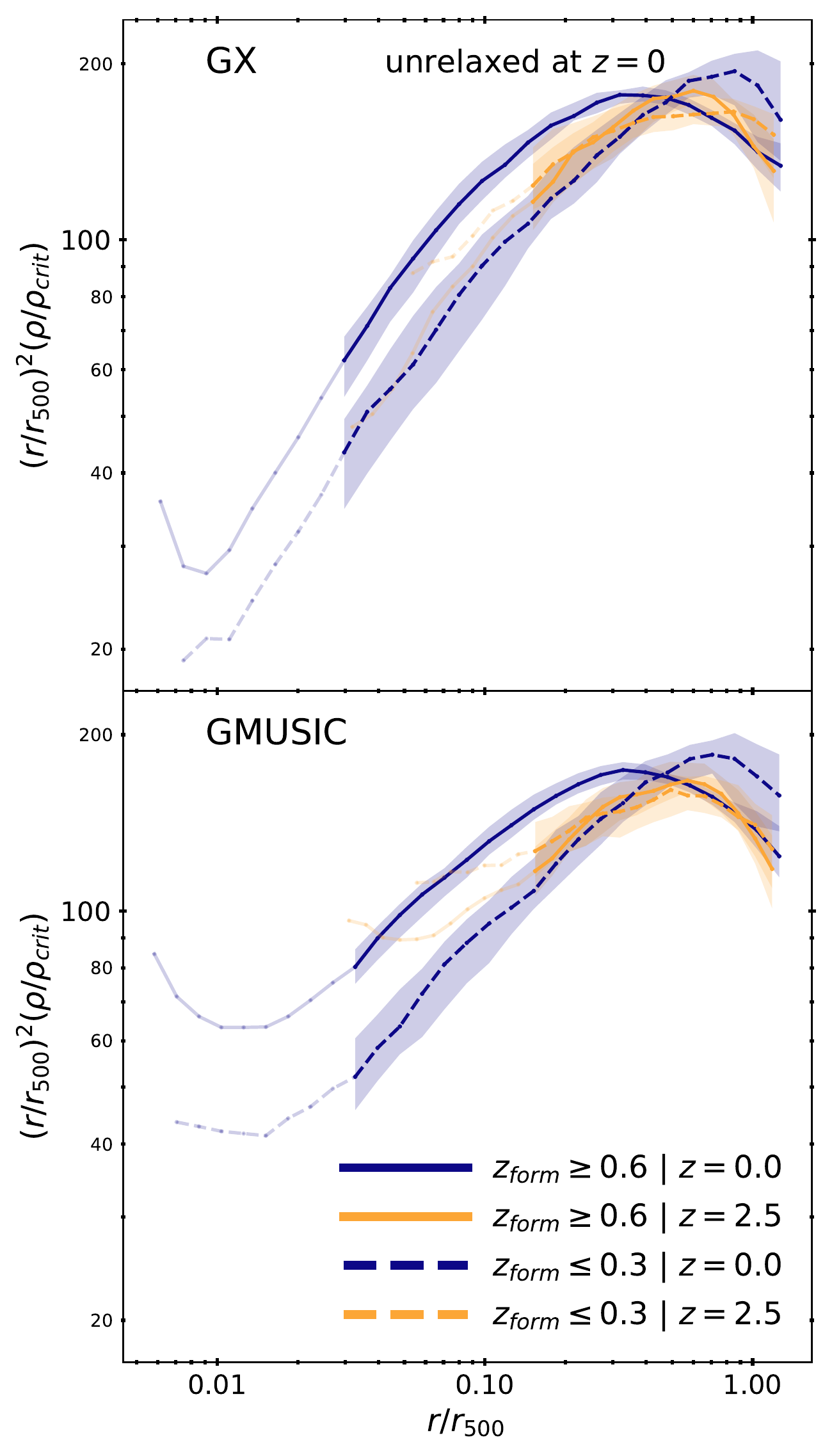}
    \caption{ The redshift evolution of the median scaled density profiles ($z=0$ and $z=2.5$) for unrelaxed early-formed ($\sim24$ per cent of the unrelaxed clusters, with $z_{\rm form} \geq 0.6$) and unrelaxed late-formed ($\sim25$ per cent, with $z_{\rm form} \leq 0.3$) clusters, from \code{GADGET-X} (top) and \code{GADGET-MUSIC} (bottom). Despite their dynamical state, a shift is observed in the early formed unrelaxed clusters.}
    \label{fig:test_shift_unrelax_lateANDearly}
\end{figure}

We now seek a better understanding of the shift in the median density profiles, as seen for the relaxed subsample. Overall, relaxed clusters formed earlier ($z_{\rm form} \sim 0.8$) than the unrelaxed ones ($z_{\rm form} \sim 0.4$). This implies that at $z=0$, as seen from the growth rates of both the physical $r_{\rm s}$ and $r_{500}$, most of the unrelaxed clusters are still in the fast accretion phase whereas the majority of the relaxed clusters have entered the slow accretion phase: contrary to the mass buildup of the unrelaxed subsample, infalling material no longer accumulates in the central region but rather in the outskirts of the clusters. Consequently, for the relaxed sample, $r_{\rm s}$ no longer grows while, along with pseudo-evolution, the infalling material induces a growth in $r_{500}$. This is in agreement with the secondary infall model \citep[e.g.][]{Lokas00,Ascasibar04,Ascasibar07} and previous numerical results \citep[e.g.][]{Bullock01,Wechsler02, Zhao03}. While this trend is observed in the median results for the relaxed subsample, we argue that the main reason why the shift is not visible in the median profiles of the whole sample (Fig. \ref{fig:density_median_relaxation}) is the scatter associated with the diversity of density profiles in the unrelaxed clusters. Since the sample is mostly dominated by unrelaxed (as classified at $z=0$), relatively late-formed ($z_{\rm form}\sim0.5$) clusters, studying the median profile evolution washes out such a shift. This can be verified in \Fig{fig:test_shift_unrelax_lateANDearly}, where we show the evolution of the median density profile of only unrelaxed clusters ($\sim272$ clusters) that formed early ($\sim24$ per cent of the unrelaxed sample, with $z\geq0.6$, solid lines) and late ($\sim25$ per cent, with $z\leq0.3$, dashed lines) in \code{G-X} (upper panel) and \code{G-MUSIC} (lower panel). We see that unrelaxed, early-formed systems also show a marginal trend for evolution of the peak position, albeit still within the respective percentiles of the unrelaxed late-formed ones. In regards to the formation time selection criteria used in the analysis, while the chosen boundary values, i.e. $z_{\rm form}=0.3$ and $z_{\rm form}=0.6$, are somewhat arbitrary, it is worth noting that picking different values, besides changing the sample sizes and the relaxed/unrelaxed fractions of clusters in each bin, will not induce drastically different results, as we can infer from \Fig{fig:test_shift_unrelax_lateANDearly}. As previously stated, the correlation between dynamical state and formation time in \Fig{fig:formation_time_distribution_relaxation} leads to the shift in the relaxed subsample, but as we showed in \Fig{fig:test_shift_unrelax_lateANDearly}, formation time dictates the shift.

\subsection{Relation to Gas Mass Profiles}\label{sec:subsec:results_gasprofiles}

\begin{figure*}
	\includegraphics[width=1\textwidth]{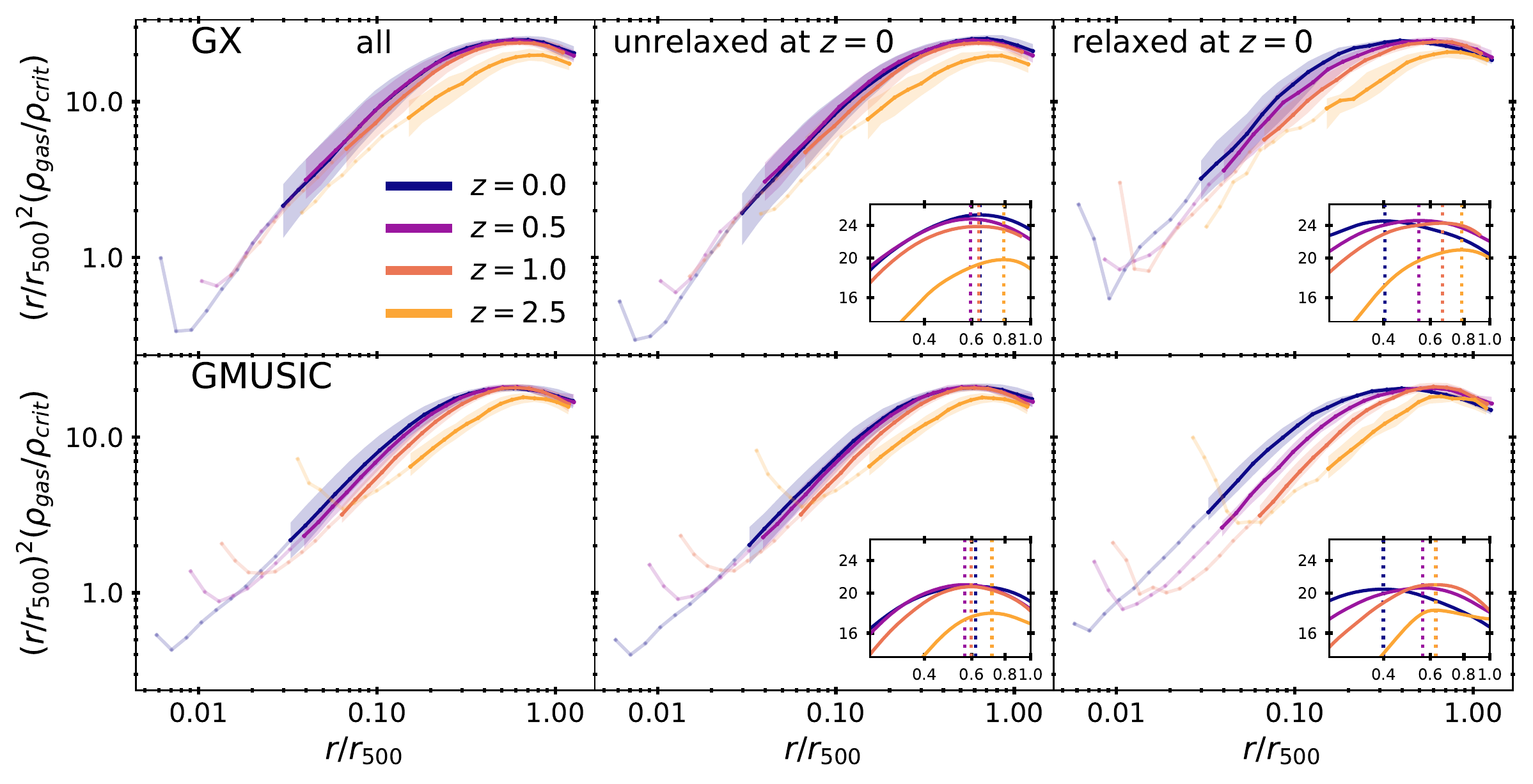}
    \caption{Median scaled gas mass density profiles of the central haloes at $z=0$ and their main progenitors at $z=0.5$, $1$, and $2.5$, classified by their dynamical state at $z=0$. The top row shows the results from \code{GADGET-X}, the bottom row from \code{GADGET-MUSIC}. The profiles show the same self-similar evolution as the total mass density profiles, as well as similar shifts once we select the relaxed subsample, up to redshift $z=1$. At $z=2.5$ we identify a slight deviation from the rest of the distribution which might be a result of the high merging activity or the star formation rate at that redshift.}
    \label{fig:density_gas_median_relaxation}
\end{figure*}

We now raise the question of how these results relate to observations. As reported by \citet{McDonald17}, the mean (hot) gas profile of massive clusters shows a pronounced self-similar evolution out to a redshift of $z\sim1.9$. Having performed full physics hydrodynamical simulations, we also have access to the gas density profile and hence show its evolution in the same manner as before in \Fig{fig:density_gas_median_relaxation} and with a focus on relaxed and unrelaxed systems. We confirm that the gas density profile follows qualitatively the self-similar evolution observed in the total mass density profile seen in Fig. \ref{fig:density_median_relaxation}: out to $z=1$, the mass-complete sample and the unrelaxed subsample evolve self-similarly, whereas the relaxed subsample shows a shift in the peak's position. Focusing in the relaxed subsample we can see that both codes evolve in a similar fashion, and that the main difference resides in the scatter in the profiles. \code{G-X}'s median profiles show more diversity due to the influence of AGN feedback (specially in the inner $r/r_{500}<0.1$ region), while \code{G-MUSIC}'s median profiles have overall less scatter at those scales. In order to present a clearer picture of the evolution of the gas profiles, we added insets for the two subsamples. As we can see in the insets, there is a broadening of the median profiles at $z=0$ compared with the profile at $z=2.5$. However, despite this broadening, there is indeed a shift in the peak's position of the relaxed subsample as the scatter at the peak's position is negligible compared with the shift between $z=0$ and $z=2.5$ (i.e. $\Delta(r/r_{500})\sim0.2-0.4$). 

By redshift $z=2.5$ the profiles deviate slightly from the rest of the distribution, which might be caused by several processes. At high redshift mergers are the main source of halo mass build-up, as they are in their early-phase accretion mode \citep[e.g.][]{Wechsler02,Burke13}. Hence, processes like ram-pressure stripping could remove the gas of the infalling satellites \citep[e.g.][]{Fujita01,Fujita04,Wang18}. The deficit could also be attributed to the high star formation rate in our simulated clusters, which peaks around $z=2.5$ \citep[as seen in Figure 4 of][]{Wang18}. Both the central galaxy and the infalling satellite galaxies from mergers produce a gas deficit at different scales. Finally, the clumpiness of gas particles due to the SPH treatment of hydrodynamics in numerical simulations \citep[e.g.][]{Hobbs13} may also be a cause, at least for \code{G-MUSIC}. This numerical issue causes the gas to clump rapidly into the halo center --contrary to the smooth accretion of gas-- and leave a lower gas density at outer halo radius.

Nevertheless, the fact that we observe a self-similar evolution of the gas density profile does not come as a surprise as the scales we are probing here are dominated by gravity and hence the gas is a biased tracer (note the different scales on the $y$-axis as compared to the previous total matter density plots). We observe that the gas profiles show inherently larger scatter, especially at small scales $(r/r_{500}<0.02)$ and for \code{G-X} (i.e. the code that features AGN feedback affecting the central regions). However, we can constrain the influence of the baryons to lie within $r/r_{500}<0.01-0.1$, depending on the redshift. Compared with observations, recent results of high redshift massive galaxy clusters show that beyond the core of the clusters ($r/r_{500}>0.3$) the level of self-similarity in gas density profiles is particular remarkable and that non-gravitational effects, such as AGN feedback, can be confined in the $r/r_{500}<0.2$ region \citep{McDonald17}. Moreover, in \citet{Ghirardini18}, the authors analyzed 12 massive ($M_{\rm 500}=3-9\times10^{14}$M$_\odot$) high-quality local ($z<1$) galaxy clusters and determined the \textit{intrinsic} scatter of thermodynamic quantities as a function of radius. The amount of scatter is minimized in the $0.2<r/r_{500}<0.8$ range, the region where the gas is highly virialized and baryonic effects are negligible. In the inner region ($r/r_{500}<0.3$) baryonic effects induce large scatter within the population, while in the outer region ($r/r_{500}\geq1$) the scatter is driven by the different accretion rates from one cluster to another \citep[see][and references therein, for more determinations of intrinsic scatter]{Eckert12,Reiprich13}.
Looking at \Fig{fig:density_gas_median_relaxation} we can see similar trends: up to $z=1$, the least amount of scatter, as described by the shaded region corresponding to the $30-70$ percentiles, is achieved for $0.3<r/r_{500}<1$. The median gas profiles from \code{G-X} follow the self-similar evolution down to $r/r_{500}\approx0.02$. However, the scatter induced by non-gravitational processes dominates in the region. On the other hand, in \code{G-MUSIC}, although deviations from the self-similar trend are visible at the same radial range, the scatter is systematically lower.

\subsection{Present day halo scales} \label{sec:subsec:present_scales}

\begin{figure*}
	\includegraphics[width=1\textwidth]{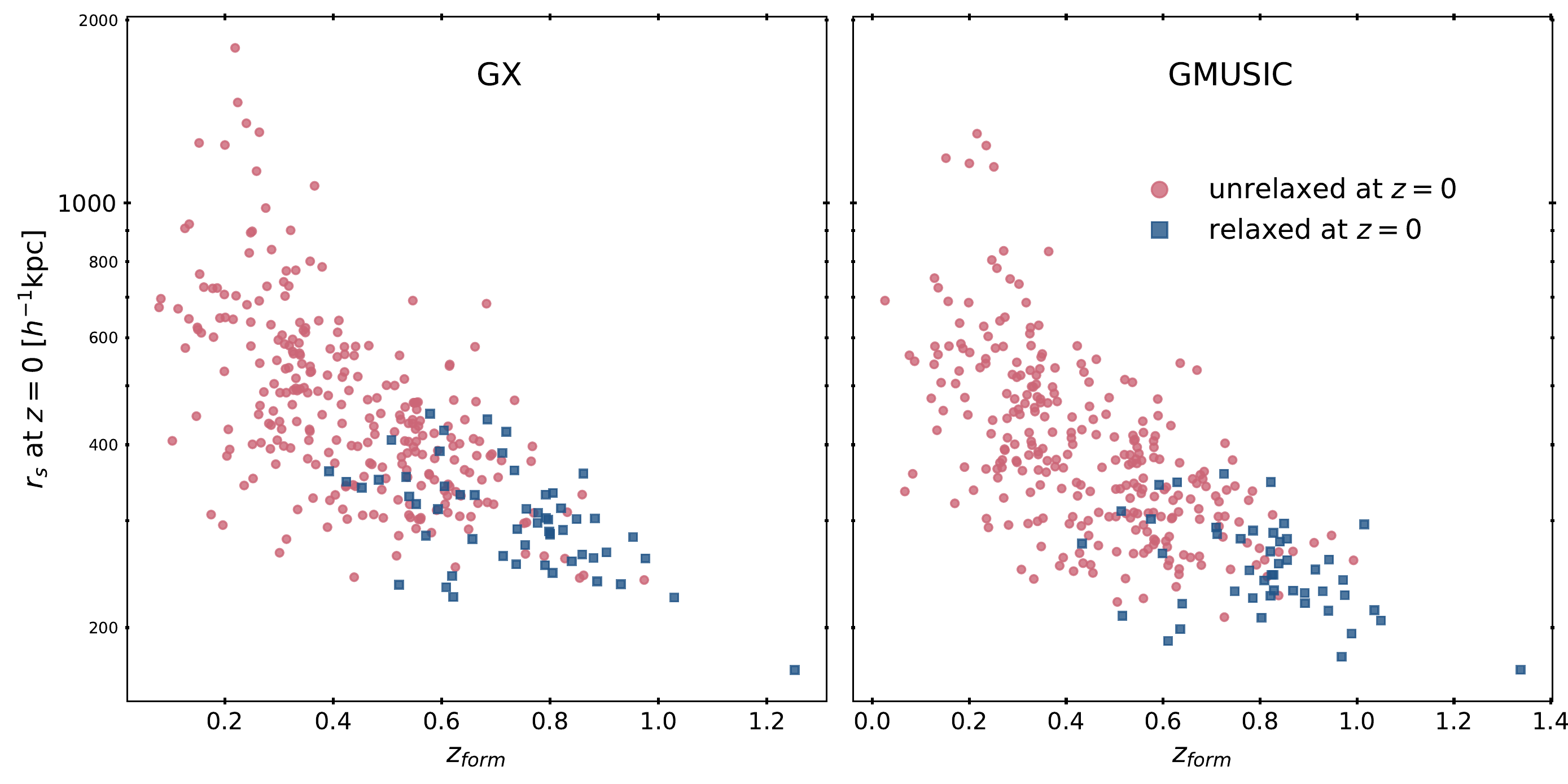}
    \caption{Scale radius $r_s$ at redshift $z=0$ correlation with formation redshift $z_{\rm form}$ for relaxed clusters (blue) and unrelaxed clusters (orange). The left panel shows the results for \code{GADGET-X} and the panel for \code{GADGET-MUSIC}. Overall, early-formed clusters ($z_{\rm form}\geq0.6$) end up having smaller values of the scale radius $r_s$ at $z=0$ ($r_s<0.3$ Mpc), while later-formed ones show a greater dispersion in the value of their scale radius, $r_s$. The unrelaxed clusters that formed at similar redshift as the relaxed ones might appear as a consequences of our relaxation classification. Since we are not tracking the dynamical state of the progenitors of the haloes but instead keep their classification at $z=0$,  processes which disrupt relaxation but do not induce a considerable change in the formation time determination (i.e. minor mergers) might produce unrelaxed clusters at $z=0$ with similar formation times as the relaxed ones. On the other hand, the unrelaxed, late-formed clusters that show a small scale radius at $z=0$ are the effect of a misidentification of the progenitor halo due to recent (major) mergers, which shows up as a sudden increase in the mass accretion history of the cluster.}
    \label{fig:rs_zform_correlation}
\end{figure*}

The main finding from the previous sub-sections (notwithstanding the dark matter-gas connection) can be summarized by \Fig{fig:rs_zform_correlation} that shows the scale radius $r_{\rm s}$ and $r_{500}$ at redshift $z=0$ for every cluster in the sample as a function of its  formation redshift $z_{\rm form}$, where we additionally colour-code the clusters by their dynamical state at $z=0$. For this plot we now require individual $r_s$ values for each cluster. These are obtained by fitting the enclosed\footnote{The enclosed mass $M(<r)$ is a cumulative \--- and generally smooth \--- profile, hence it is easier to obtain a good fit. } radial mass distribution $M(<r)$ to the functional form of a NFW profile \citep[e.g.][]{Lokas01}: 
\begin{align}
\frac{M(s)}{M_{200}} = g(c)\left(\ln{(1+cs)}-\frac{cs}{1+cs}\right) \label{eq:nfw_enclosed_mass}
\end{align}
with
\begin{align}
s    &= \frac{r}{r_{200}},\\
c    &= \frac{r_{200}}{r_s},\\
g(c) &= \frac{1}{\ln{(1+cs)} - cs/(1+cs)}
\end{align}

\noindent
where $c$, the only free parameter, is the concentration of the cluster; $r_s$ is now simply calculated as
\begin{equation}
r_s = \frac{r_{200}}{c} .
\end{equation}

The overall trend shows that relaxed clusters end up with smaller values of $r_s$ at $z=0$ than the unrelaxed ones. It is worth noting that, as mentioned previously, there are unrelaxed clusters which formed at similar redshift as relaxed ones. Morever, the early-formed unrelaxed haloes show a considerable amount of scatter in their scale radius $r_{s}$ at $z=0$, there are unrelaxed haloes with comparable $r_s$ values to the relaxed ones that formed earlier. These effects are mainly due to two issues. First, the classification of the mass-complete sample by its dynamical state is done at redshift $z=0$ and it is kept up to $z=2.5$, i.e. we do not track information about the dynamical state of the progenitors of the haloes. A cluster classified as relaxed at $z=0$ might not pass our relaxation criteria at higher redshifts; and vice versa, an unrelaxed cluster at $z=0$ might have been a relaxed cluster at some point in its past and only recently being disturbed again. If an initially relaxed cluster becomes unrelaxed without undergoing a significant change in its formation time determination, i.e. a change in its subhalo fraction $f_{s}$ due to minor mergers, it would show as an unrelaxed halo with similar formation times and $r_s$ values at $z=0$ to the relaxed haloes. Second, a recent (major) merger might induce a steep mass accretion close to $z=0$, which --given our formation time definition-- consequently translates into a lower than expected value for $z_{\rm form}$. Thus, we could understand the unrelaxed haloes with comparable $r_s$ values to the relaxed ones as haloes that in fact formed earlier (i.e. $z>0.3$) and have been disturbed recently by a major merger and, as such, they have a lower $r_{s}$ value than the rest of the haloes in the early-formed bin.

\section{Conclusions} \label{sec:conclusions}

In this work we used the mass-complete sample of the \textsc{The Three Hundred} data set \citep{Cui18}, consisting of 324 galaxy cluster haloes with median mass $M_{200} = 8.2\times h^{-1}10^{14}$M$_{\odot}$. For cluster-sized haloes, self-similarity is expected to dominate the redshift evolution of their mass density profiles, although baryonic non-gravitational interactions are known to disrupt the effect. We have used our data set to verify this.

We tracked the evolution of the central haloes found at $z=0$ back to $z=2.5$. We found that the density profiles of the whole sample is consistent with a self-similar evolution, suggesting that the density profile is stable and already in place even at $z\geq 2$, long before their formation time and as found in dark matter only simulations by other workers in the field (e.g. LB+18). However, when separating the sample according to their dynamical state at $z=0$, the relaxed clusters ($\sim16$ per cent of the total sample) show a shift in where their scaled median density profile peaks. We found that this shift is a result of a different evolution of $r_{s}$, the scale radius of a NFW density profile, and $r_{500}$ between the unrelaxed and relaxed subsamples. This is understood within the context of a two-phase accretion model for halo mass growth, as theoretically argued \citep[e.g.][]{Gunn72,Gunn77,Ascasibar04} and found in numerical simulations \citep[e.g.][]{Wechsler02, Zhao03}. While unrelaxed clusters are still in their fast accretion mode at $z<0.5$, in which infalling material is accreted to the cluster centre inducing rapid growth in both $r_{\rm s}$ and $r_{500}$; relaxed clusters reached the slow accretion phase at $z<0.5$ and thus infalling material remains in the outer region of the cluster while the inner region evolves almost unperturbed. Therefore, $r_{\rm s}$ slows its growth at the later stages of the cluster's evolution while $r_{500}$ keeps growing due to the effects of infalling material (and also pseudo-evolution) at the outskirts of the halo. To test this hypothesis we classified the sample by their formation time, defined as the redshift at which the fitted mass growth curve of the cluster reaches $50$ per cent of its total mass $M_{200}$ at $z=0$, regardless of their dynamical state. We find that once binned by their formation time, the same trend is observed. Early formed ($z_{\rm form}\geq 0.6$) clusters end up with lower values of $r_{\rm s}/r_{500}$ than late-forming clusters ($z_{\rm form}\leq 0.3$). Studying the individual evolution of $r_{\rm s}$ and $r_{500}$ we observe the same individual evolution as in the dynamical state classified samples. We argue that this shift, although potentially visible also for unrelaxed haloes due to the overlap between unrelaxed early-formed haloes and relaxed late-formed ones, is not visible in the median profiles of unrelaxed profiles due to the scatter associated with their diversity of density profiles. Relaxed clusters, on the other hand, have stable density profiles, thus less scatter, and consequently the shift is visible in this subsample. 

We also analyzed the gas profiles of the mass-complete sample. We find that the gas profile follows qualitatively the self-similar evolution of the total density profile with only a considerable deviation at $z=2.5$ which can be attributed to the combined effect of the merger activity and star formation rate at that redshift, and numerical effects from SPH. This follows naturally from the fact that gravity is the main driving force at the scales studied here and thus gas is a biased tracer of the underlying dark-matter distribution of the cluster. Even the powerful AGN in the centre of the clusters in the \code{GADGET-X} simulations does not affect the density profile at distances $r\gtrsim 0.1r_{500}$, although it increases the scatter in the inner regions of the cluster. The influence of the baryons on the total (dark matter and baryons) density profile can be safely constrained within the innermost part of the cluster, at $r/r_{500}<0.01-0.1$, depending on the redshift. We note that our results are in agreement with observations \citep[e.g.][]{McDonald17,Ghirardini18}, although the shift observed in our dynamically relaxed/early-formed subsample is not detected in observational studies due to the implicit selection of the sample in every observation.

The implications of this work form something of a warning to those seeking to use massive galaxy clusters as cosmological probes. The significant shuffling of the rank order in mass of massive clusters should not be a surprise: at the very massive end the mass function is steep and small changes in mass can make a large difference. Also, in the era of league tables those close to the top know full well that the likely direction of travel can only be down. This fact of life as a giant galaxy cluster makes it very easy to select a biased sample and both theorists and observers need to be aware of the consequences of this. The most massive clusters today are not the same clusters as those most massive at $z=1$. Selecting a mass-limited sample, such as that used by LB+18 is one way out of this connudrum, but such a choice may obscure certain theoretical ideas about cluster growth. Certainly great care needs to be expressed in sample selection as it is clear that without a detailed knowledge of completeness bias may be introduced into any measure of evolution. We have also demonstrated that care should be taken when selecting a relaxed sample of objects: these are the set where is it easiest to see evolution and growth along theoretical lines because this set have lower scatter in their properties. As such, they may not actually evolve self-similarly.
\section*{Acknowledgements}

The work has received financial support from the European Union's Horizon 2020 Research and Innovation programme under the Marie Sklodowskaw-Curie grant agreement number 734374, i.e. the LACEGAL project\footnote{\url{https://cordis.europa.eu/project/rcn/207630\_en.html}}. The workshop where this work has been finished was sponsored in part by the Higgs Centre for Theoretical Physics at the University of Edinburgh.

RM, AK, WC, and GY are supported by the {\it Ministerio de Econom\'ia y Competitividad} and the {\it Fondo Europeo de Desarrollo Regional} (MINECO/FEDER, UE) in Spain through grant AYA2015-63810-P. AK is also supported by the Spanish Red Consolider MultiDark FPA2017-90566-REDC and further thanks The Perishers for let there be morning. CP acknowledges the Australia Research Council (ARC) Centre of Excellence (CoE) ASTRO 3D through project number CE170100013. 
The authors would like to thank The Red Espa\~nola de Supercomputaci\'on for granting us computing time at the MareNostrum Supercomputer of the BSC-CNS where most of the cluster simulations have been performed. Part of the computations with \code{GADGET-X} have also been performed at the `Leibniz-Rechenzentrum' with CPU time assigned to the Project `pr83li'.

We benefited from valuable discussions with Yago Ascasibar. We would also like to thank the anonymous referee for their thorough comments that helped to improve the paper.


\bibliographystyle{mnras}
\bibliography{archive}



\appendix

\section{Median profiles properties}\label{app:prof_properties}In \Tab{tab:prof_properties} we present the properties of the median profiles: the median $r_{500}$, the validation radius $r_{\rm valid}$ (which is equal to the maximum convergence radius in the MDPL2 simulation $r^{\rm MDPL2}_{\rm conv}$), the inner and outer limits of the region where at least 50 per cent of the sample contributes to the median $[r^{50\%}_{\rm in},r^{50\%}_{\rm in}]$ (i.e. the threshold used for our analysis), and the inner and outer limits of the region where \textit{all} the clusters contribute to the median $[r^{100\%}_{\rm in}, r^{100\%}_{\rm out}]$. Note that the range of interest (i.e. about the peak position) always resides inside the region where \textit{all} clusters contribute to the median ($x_{\rm peak}\sim 0.3-0.8 \in [r^{100\%}_{\rm in}, r^{100\%}_{\rm out}]/(r_{500})$ for every redshift). Moreover, up to $z<2.5$ our `50 per cent criterion' only affects the outer median profile and \textit{not} the inner one due to the validation radius we employed (i.e. $r_{\rm valid} > r^{100\%}_{\rm in}$). At higher redshift ($z\geq2.5$) the validation radius does affect the inner limit. Since \code{AHF} uses adaptive binning based on the number of particles of a halo, at higher redshift the $[r^{100\%}_{\rm in}, r^{100\%}_{\rm out}]$ region becomes smaller. Nevertheless, the peak position $x_{\rm peak}$ is always well inside the $100\%$ region, thus, its determination is \textit{insensitive} to the choice of the inner and outer limits threshold.

\begin{table*}
\centering
\caption{Properties of the median profiles of the whole sample: the median $r_{500}$, the validation radius $r_{\rm valid}$ (which is equal to the maximum convergence radius in the MDPL2 simulation $r^{\rm MDPL2}_{\rm conv}$), the inner and outer limits of the region where at least 50 per cent of the sample contributes to the median $[r^{50\%}_{\rm in},r^{50\%}_{\rm in}]$, and the inner and outer limits of the region where \textit{all} the clusters contribute to the median $[r^{100\%}_{\rm in}, r^{100\%}_{\rm out}]$.  All the values are in units of (comoving) $h^{-1}$ kpc.} 
\label{tab:prof_properties}
\resizebox{\textwidth}{!}{
\begin{tabular}{lccccccccccc}
\hline
\hline
Redshift & \multicolumn{2}{c}{$r_{500}$} & $r_{\rm valid}$ & \multicolumn{2}{c}{$r^{50\%}_{\rm in}$} & \multicolumn{2}{c}{$r^{50\%}_{\rm out}$} & \multicolumn{2}{c}{$r^{100\%}_{\rm in}$}    & \multicolumn{2}{c}{$r^{100\%}_{\rm out}$} \\
& \code{G-X} & \code{G-MUSIC} & & \code{G-X} & \code{G-MUSIC} & \code{G-X} & \code{G-MUSIC} & \code{G-X} & \code{G-MUSIC} & \code{G-X} & \code{G-MUSIC}\\
\hline       
$z=0$       & 1005.9 & 1009.5    & 28       & 6.2  & 5.9    & 1279.0 & 1268.4        & 16.6  & 10.4     & 861.5 & 1046.8 \\
\hline
$z=0.5$     & 979.2  & 980.6     & 37       & 10.2 & 8.8    & 1243.0 & 1245.3        & 26.7 & 26.5      & 846.4 & 1036.8 \\
\hline
$z=1$       & 807.3  & 824.9     & 44       & 12.3 & 11.1   & 889.9  & 983.0         & 31.1 & 37.2      & 889.8 & 696.4  \\
\hline
$z=2.5$     & 398.5  & 409.7     & 55       & 15.1 & 14.7   & 477.5  & 484.2         & 71.6  & 97.7     & 338.2 & 361.9  \\
\hline
\hline
\end{tabular}
}
\end{table*}

\section{Most massive haloes at each redshift}\label{app:most_massive}
The analysis presented in the main body of the paper traced a mass-complete sample, i.e. the central haloes of the 324 simulated $15h^{-1}$ Mpc regions, from redshift $z=0$ backwards in time. While this allows us to directly measure and quantify the \emph{evolution} of density profiles, the progenitors identified at any higher redshift certainly do not form a mass-complete sample anymore. This is why, for instance, LB+18 have chosen the alternative approach of always using the most massive clusters at each redshift studied. In order to test any differences entering the analysis due to these varying strategies, we have also restricted our analysis to the 25 most massive clusters at $z=0$,$0.5$, and $1.0$; we confirm that they do in fact form a mass-complete sample at each of these redshifts \citep[cf.][]{Cui18}. In this way we extract a sample similar to that studied by LB+18. In Tab. \ref{tab:stat_values_halo_mass_massiveselect} we show the minimum, median, and maximum $M_{200}$ values for the 25 most massive cluster sample for both \code{GADGET-X} and \code{GADGET-MUSIC}. 

\begin{table}
\centering
\caption{Minimum, median, and maximum mass values of the 25 most massive haloes at each redshift (see text for further details). In each row, the left value corresponds to the \code{GADGET-X} simulation, the right one to \code{GADGET-MUSIC}. All values are in units of $10^{14}h^{-1}$M$_{\odot}$.}
\label{tab:stat_values_halo_mass_massiveselect}
\resizebox{\columnwidth}{!}{
\begin{tabular}{lcccccc}
\hline
\hline
Redshift & \multicolumn{2}{c}{min($M_{200}$)}    & \multicolumn{2}{c}{med($M_{200}$)}  & \multicolumn{2}{c}{max($M_{200}$)}\\
   & \code{G-X} & \code{G-MUSIC} & \code{G-X} & \code{G-MUSIC} & \code{G-X} & \code{G-MUSIC}\\
\hline
$z=0$      & 13.57 & 13.81                         & 15.76 & 15.68                      & 26.21 & 26.22 \\ 
\hline
$z=0.5$    & 7.43  & 7.06                          & 8.44  & 8.30                       & 18.93 & 18.95 \\ 
\hline
$z=1$      & 3.68  & 3.78                          & 4.16  & 4.23                       & 6.90  & 6.89  \\ 
\hline
\hline
\end{tabular}
}
\end{table}

\begin{figure*}
	\includegraphics[width=1\textwidth]{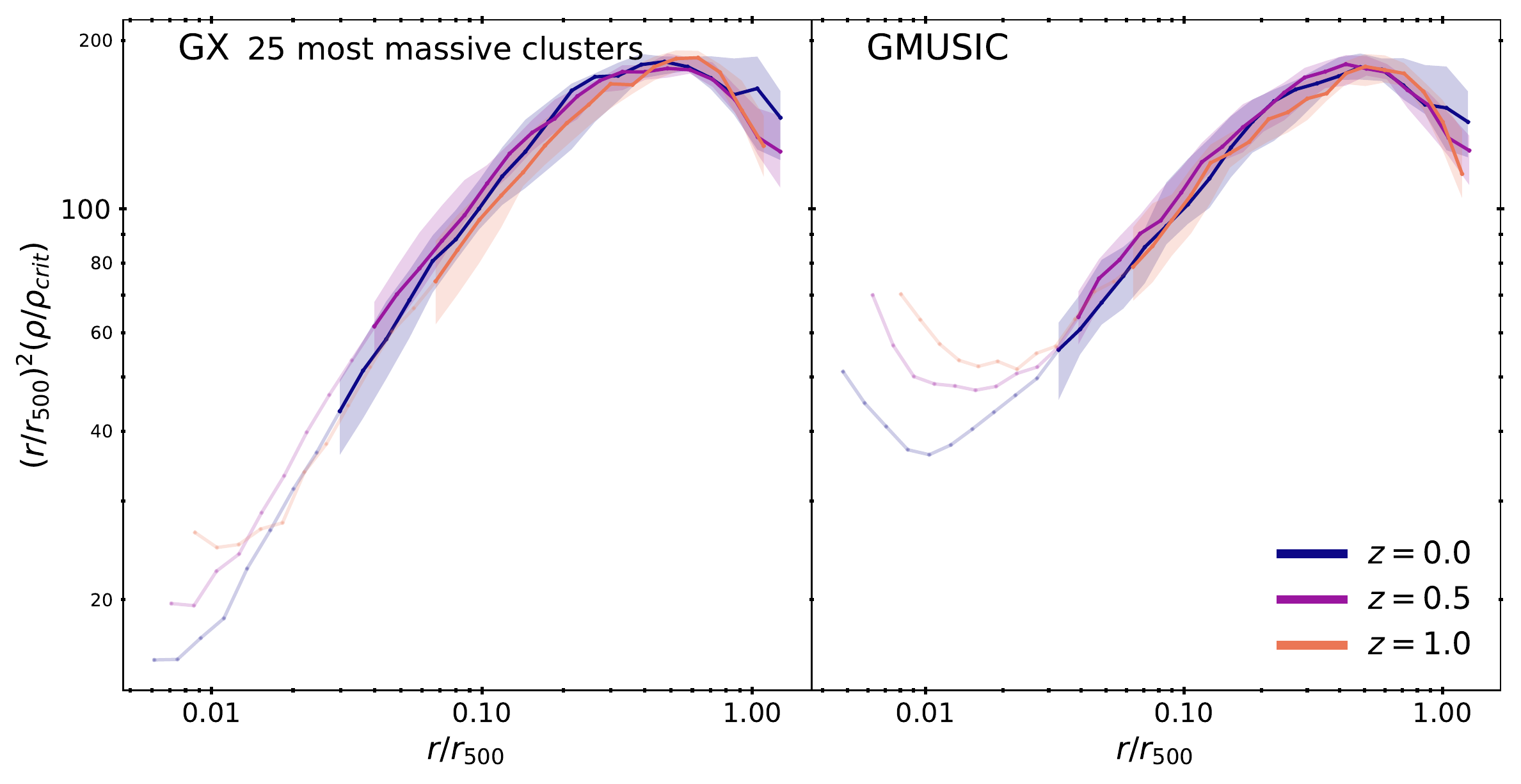}
    \caption{Median scaled mass density profiles for the 25 most massive galaxy clusters at redshift $z=0$, $0.5$, and $1$ for the two hydrodynamical simulations in the sample, \code{GADGET-X} (left) and \code{GADGET-MUSIC} (right). The shaded region shows the 30-70 percentiles.  Unvalidated values are shown in lighter colors. The baryonic component of the clusters is reflected as a turnaround in the innermost region, at $r/r_{500}\sim 0.02$  and $r/r_{500}\sim 0.03$ at $z=1$ for \code{GADGET-X} and \code{GADGET-MUSIC}, respectively. Although \code{GADGET-X} has implemented a model for AGN feedback, the self-similarity evolution is preserved.}
    \label{fig:density_median_mostmassive}
\end{figure*}

In Fig. \ref{fig:density_median_mostmassive} we show the median scaled mass density profiles for the 25 most massive galaxy clusters at redshift $z=0$, $0.5$, and $z=1$. In agreement with the results of LB+18, the mass distribution of massive galaxy clusters is in place at $z>1$, despite the presence of non-gravitational radiative baryonic processes. Note that the baryonic turnaround in the profiles is more pronounced for \code{GADGET-MUSIC}. This is mainly attributed to an over-production of stars due to the lack of AGN feedback in the code. However, as we can see from the \code{GADGET-X} results, the AGN feedback is not strong enough to disrupt the total mass distribution considerably, thus the self-similarity of the profiles is preserved down to $r/r_{500}\approx 0.03$.

\section{Dark matter only simulation}\label{app:dm_only}
\begin{figure*}
	\includegraphics[width=1\textwidth]{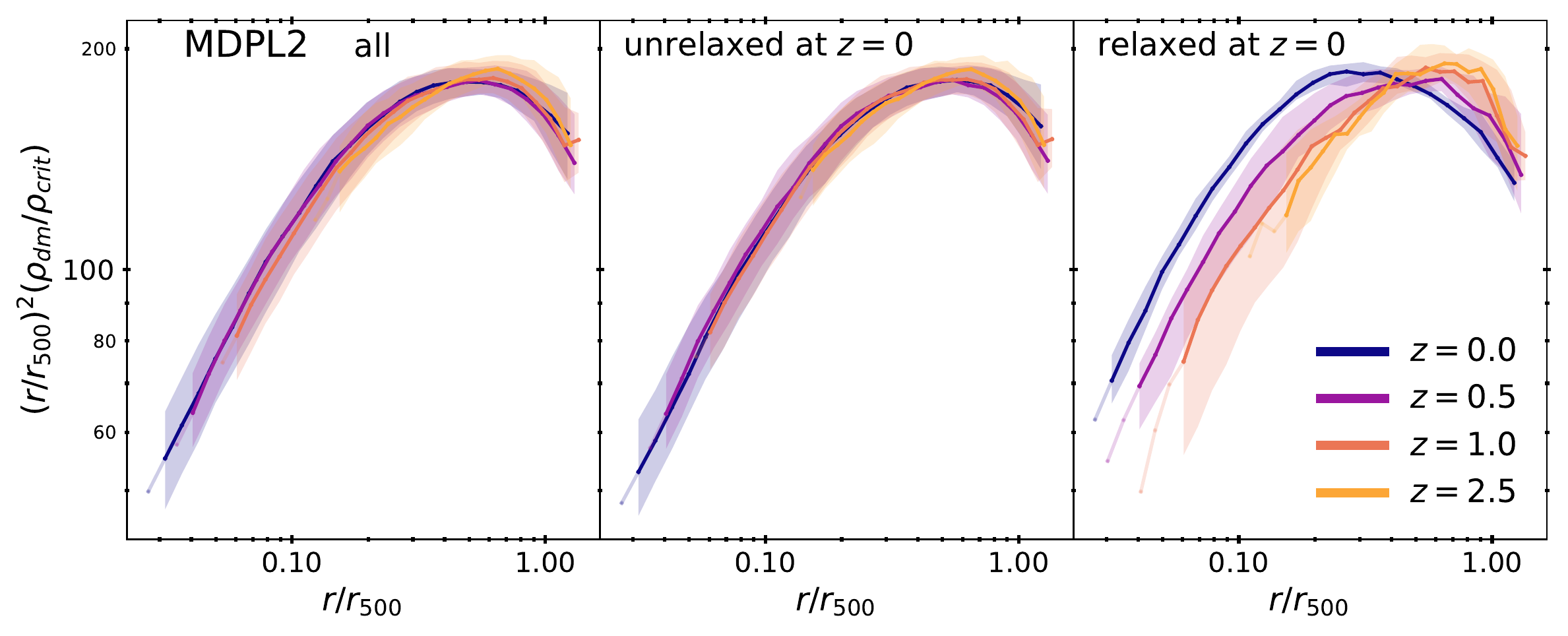}
    \caption{Median scaled mass density profiles for 324 dark-matter-only central haloes from the MDPL2 simulation at redshift $z=0$, $0.5$, $1$ and $2.5$. Once classified by their dynamical state at $z=0$, the relaxed subsample ($\sim11$ per cent of the total sample, right column) presents the same $r_{s}$ evolution as the results from the hydrodynamical simulations.}
    \label{fig:density_median_MDPL2_relaxation}
\end{figure*}

In order to directly quantify any influence on our results stemming from the modelled baryonic physics, we have repeated all our analysis for the analogues of our clusters as found in the dark matter only MDPL2 simulation. Proceeding in a similar fashion, once classified by their dynamical state at $z=0$, a similar shift is observed for the relaxed subsample ($\sim11$ per cent of the total sample) when tracking their progenitor's evolution-- as can be verified in \Fig{fig:density_median_MDPL2_relaxation}. The baryonic influence can be safely attributed to the innermost region of the cluster; it does not affect the results presented here and the conclusions drawn, respectively.

\bsp	
\label{lastpage}
\end{document}